\documentclass[letterpaper,11pt]{article}
\usepackage{amsmath}
\usepackage{amsfonts}
\usepackage{amssymb}
\usepackage{mathrsfs}
\usepackage{multicol}
\usepackage{color}
\usepackage{xcolor}
\usepackage{graphicx}
\usepackage{pstricks}
\usepackage{pst-node}
\usepackage{wrapfig}
\usepackage{enumerate}
\usepackage{caption}
\usepackage[toc,page]{appendix}
\usepackage{cite}
\usepackage{subfigure}
\usepackage{amsfonts,amssymb}
\usepackage{amsthm}
\usepackage{amscd}
\definecolor{gray1}{gray}{0.1}
\definecolor{gray2}{gray}{0.2}
\definecolor{gray3}{gray}{0.3}
\definecolor{gray4}{gray}{0.4}
\definecolor{gray5}{gray}{0.5}
\definecolor{gray6}{gray}{0.6}
\definecolor{gray7}{gray}{0.7}
\definecolor{gray8}{gray}{0.8}
\definecolor{gray9}{gray}{0.9}

\newpsobject{showgrid}{psgrid}{subgriddiv=1,griddots=10,gridlabels=6pt}

\definecolor{dark-green}{rgb}{0,0.7,0}
\definecolor{dark-blue}{rgb}{0,0.2,0.5}
\definecolor{med-blue}{rgb}{0,0.7,1}
\definecolor{mblue}{rgb}{0,0.2,1}
\definecolor{cnc}{rgb}{0.8,0,0}
\definecolor{light-red}{rgb}{1,0.8,0.8}
\definecolor{dark-yelow}{rgb}{1,0.8,0}
\definecolor{light-blue}{rgb}{0.8,0.9,1}
\definecolor{verylight-blue}{rgb}{0.93,0.95,1}
\definecolor{light-yelow}{rgb}{1,0.9,0.8}
\definecolor{grey}{gray}{0.88}

\newpsobject{showgrid}{psgrid}{subgriddiv=1,griddots=10,gridlabels=6pt}

\usepackage[normalem]{ulem}

\begin{document}

\thispagestyle{empty}

\setlength{\abovecaptionskip}{10pt}

\begin{center}
{\Large\bfseries\sffamily{Geodesic Motion in a Swirling Universe: \\[1ex]
The complete set of solutions}}
\end{center}
\vskip 1cm

\begin{center}
{
\bfseries{\sffamily{Rog\'erio Capobianco$^{\rm 1}$}},
\bfseries{\sffamily{Betti Hartmann$^{\rm 2}$}}, and
\bfseries{\sffamily{Jutta Kunz$^{\rm 3}$}}
}\\
\vskip 0.5cm

{$^{\rm 1}$\normalsize{Instituto de F\'isica de S\~ao Carlos, Universidade de S\~ao Paulo, S\~ao Carlos, S\~ao Paulo 13560-970, Brazil}}\\
\vskip 0.1cm

{$^{\rm 2}$\normalsize{Department of Mathematics, University College London, Gower Street, London, WC1E 6BT, UK }}\\
\vskip 0.1cm
{$^{\rm 3}$\normalsize{Institut f\" ur Physik, Carl-von-Ossietzky Universit\"at Oldenburg, 26111 Oldenburg, Germany}}\\
\end{center}

%\date{\today}

\vspace{1cm}

\begin{abstract} 
    We study the geodesic motion in a space-time describing a swirling universe. 
    We show that the geodesic equations can be fully decoupled in the Hamilton-Jacobi formalism leading to an additional constant of motion. 
    The analytical solutions to the geodesic equations can be given in terms of elementary and elliptic functions. 
    We also consider a space-time describing a static black hole immersed in a swirling universe. In this case, full separation of variables is not possible and 
    the geodesic equations have to be solved numerically. 
\end{abstract}
 
\vspace{1cm}

\centerline{\today}
 
%%%%%%%%%%%%%%%%%%%%%%%%%%%%%%%%%%%%%%%%%%%%%%%%%%%%%%%%%%%%%%%%%%%%%%%%%%%%%%
\section{Introduction}

General Relativity (GR) is a geometric approach to gravity proposed by Einstein in 1915 \cite{einstein}. 
Ever since, GR has proved to be a highly successful theory, being able to stand the so-called \textit{classical tests of GR}, such as the advance of Mercury's perihelion and the bending of light by a gravitational source \cite{Will:2005va}. 
In recent times, it has again been successfully tested by the direct observation of gravitational waves by the LIGO/Virgo collaboration \cite{LIGOScientific:2016aoc} as well as the first image of a black hole by the Event Horizon Telescope collaboration \cite{event2019first}.

Mathematically, GR results in solving coupled, nonlinear partial differential equations that describe the coupling of the curvature of space-time to the energy-momentum content of the space-time. 
These are called the Einstein field equations.  
A general derivation of solutions of the field equations is a formidable task. 
However, when space-time symmetries are imposed, the field equations can sometimes be solved ana\-lytically, leading to 
scientifically relevant settings. 
An important example is the Kerr family of solutions, which describes the gravitational field 
of asymptotically flat rotating vacuum black holes \cite{kerr}.  
In this case, the space-time is stationary and axially symmetric.

In 1968 Ernst proposed a method to obtain  
axially symmetric solutions of the Einstein field equations \cite{ernst1968new,ernst1968new2}. 
In this approach, the field equations are replaced by an equation -- the so-called {\it Ernst equation} -- for a complex-valued gravitational potential, often referred to as \textit{Ernst potential}. 
The Ernst equation is invariant under a set of transformations. 
This can be used to construct new solutions starting from a \textit{seed solution} (see e.g.~\cite{astorino2020enhanced} for an introduction to this so-called {\it Ernst generating technique}). 
Among the above mentioned transformations, the Harrison transformation provides a mechanism to immerse seed solutions, e.g.~black holes, into non-trivial backgrounds, such as the Melvin magnetic universe \cite{melvin1964pure,ernst1976black,gibbons2013ergoregions}. 
Recently, the Ernst formalism has been used to derive a new solution, describing black holes in a so-called swirling universe \cite{astorino2022black} by using the Ehlers' transformation.  

In this paper, we investigate the geodesic structure of this swirling universe space-time. 
Geodesics are nothing else but the equations of motion for a particle in free fall in a given geometry. 
The solutions of these equations are crucial in understanding the structure of the space-time. 
Often, finding explicit solutions to the geodesic equations is impossible, but in some cases this has been achieved.  
An example are the solutions of the geodesic equations in the Schwarzschild space-time in terms of the Weierstrass $\wp$-function \cite{hagihara1930theory}.
For the Kerr solution it was shown by Carter \cite{carter1968global} that the geodesic equations can be completely decoupled when employing the Hamilton-Jacobi approach. 
A new constant of motion, the Carter constant, appears, when separating the radial and polar angular motion. 
The underlying reason is the existence of Killing and Killing–Yano tensors for vacuum Petrov type D solutions in the absence of acceleration \cite{Demianski:1980mgt,Frolov:2008jr}. 
In particular, the approach can be used to describe the motion of charged particles in the Reissner-Nordstr\"om space-time \cite{grunau2011geodesics} and the Kerr-Newman space-time \cite{hackmann2013charged}, where particles no longer move on geodesics. 

This paper is organized as follows: in Section~\ref{The swirling universe} we review the swirling universe solution and derive the geodesic equations. 
In Section~\ref{solutions} we present the complete set of solutions to the geodesic equations. 
We show examples of orbits for massless and massive particles, respectively, in Section~\ref{Orbits}. 
In Section~\ref{BHswirling}, we present some results for the geodesic motion in a space-time describing a Schwarzschild black hole immersed in a swirling universe. 
We conclude with a summary and outlook.

\section{Geodesics in a swirling universe}
\label{The swirling universe}

\begin{figure}[h]
    \centering
    \includegraphics[scale=.6]{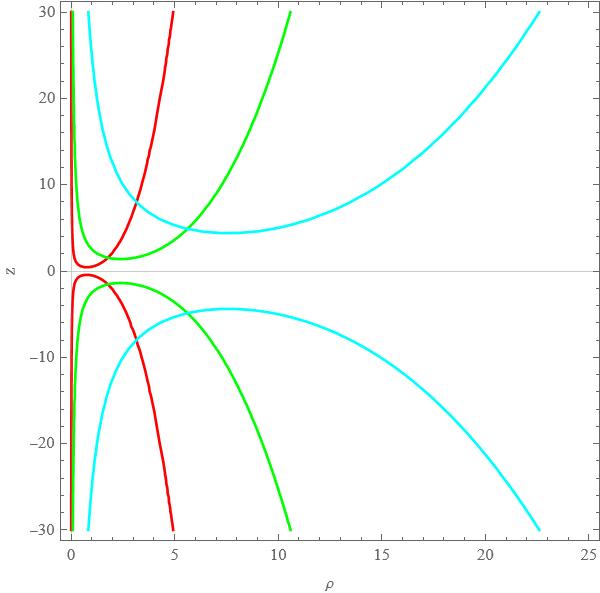}
    \includegraphics[scale=.6]{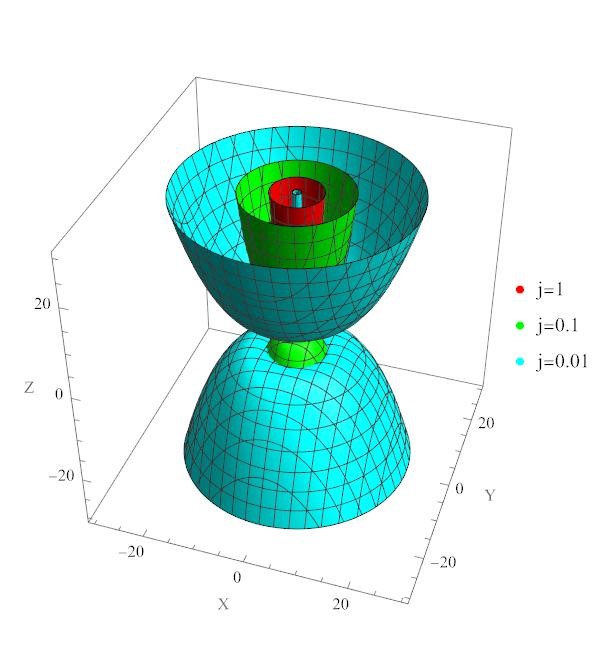}
    \caption{The ergoregion of the swirling universe solution for different values of the parameter $j$ in the $z$-$\rho$-plane (left) and rotated around the symmetry axis (right).}
    \label{figure 1 - Erogregions background}
\end{figure}

The swirling universe solution is the rotating background solution discussed in \cite{astorino2022black} and first presented in \cite{Carter:1968ks}.
This solution can be constructed by using the Ernst formalism, in particular, by applying the Ehlers' transformation to a stationary and axisymmetric seed solution. 
This transformation then embeds the seed into a rotating background, the swirling universe. 
In cylindrical coordinates $(t,\rho,\phi,z)$ the metric tensor of the swirling universe reads \cite{astorino2022black}~:
\begin{equation}\label{eq:background}
    {\rm d}s^2 = F(\rho) \left( -{\rm d}t^2 + {\rm d}\rho^2 + {\rm d}z^2 \right) + \frac{\rho^2}{F(\rho)} \left( {\rm d}\phi + \omega(z) {\rm d}t \right)^2 \ ,
\end{equation}
where $F(\rho)=1+j^2 \rho^4$ and $\omega(z) = 4jz $. 
This metric belongs to the Petrov type D class. 
This space-time possesses an ergoregion when \cite{gibbons2013ergoregions}
\begin{equation}
\label{eq: ergoregion}
    |4jz\rho| > 1 + j^2 \rho^4 \ .
\end{equation}
To better understand the structure of the space-time, we present the ergoregion for three different values of the swirling parameter $j$ in Fig.~\ref{figure 1 - Erogregions background}. 
This ergoregion is not compact, unlike the case of the Kerr space-time, but extends infinitely in $z$-direction.

So far no detailed quantitative analysis of the geodesic equations in this space-time has been given.
This is what we will do here. \\

The motion of a test particle in free fall in a given space-time is given by~:
\begin{equation}
\label{geodesics: geodesic equation}
    \frac{{\rm D}^2 x^\lambda}{{\rm D}\tau^2} = \Ddot{x}^\lambda + \Gamma^{\lambda}_{\mu \nu} \dot{x}^\mu \dot{x}^\nu = 0 \ ,
\end{equation}
where ${\rm D}/{\rm D}\tau$ denotes the covariant derivative with respect to the affine parameter $\tau$, and the dot denotes an ordinary derivative with respect to $\tau$.
In $(3+1)$-dimensional space-time, this equation leads to four coupled non-linear ordinary differential equations.  
Equivalent formulations of (\ref{geodesics: geodesic equation}) can be given using the Lagrangian, Hamiltonian and the Hamilton-Jacobi formulation, respectively. 

For the swirling universe, two cyclic variables can be directly found from the space-time symmetries. 
Since this space-time is stationary and axially symmetric, the two constants of motion are related to the conservation of a particle's total energy and angular momentum about the symmetry axis, respectively. 
These read~:
\begin{equation}
\label{geodesics: conserved quantities cyclic variables}
   -E:=p_{t}= g_{tt}\dot{t} + g_{t\phi}\dot{\phi} \ \ , \ \ \qquad L:=p_{\phi}=g_{\phi\phi}\dot{\phi} + g_{t\phi} \dot{t} \ .
\end{equation}
In addition, the normalization condition gives rise to a third constant of motion and is related to the conservation of the particle's rest mass~:
\begin{equation}
\label{geodesics: conserved quantities normalization}
    g_{\mu \nu} \dot{x}^\mu \dot{x}^\nu = \chi \ ,
\end{equation}
where $\chi=-1$ for time-like orbits and $\chi=0$ for light-like orbits, respectively.

Solving (\ref{geodesics: conserved quantities cyclic variables}) for $\dot{t}$ and $\dot{\phi}$ we then find~:
\begin{eqnarray}
\label{geodesics: conserved quantities E and L}
    F \dot{t} &=& E+ 4jLz \ ,   \nonumber \\ 
    \hspace{2cm} 
    F \dot{\phi} &=& \frac{L}{\rho^2}F^2 -4jz \left(E+ 4jLz \right) \ .
\end{eqnarray}
A fourth constant of motion can be obtained by making use of the separability of the Hamilton-Jacobi equation:
\begin{equation}
\label{geodesics: Hamilton-Jacobi equation}
    2\frac{\partial S}{\partial \tau} = g^{\mu\nu} \left(\partial_\mu S \right) \left(\partial_\nu S \right) , 
\end{equation}
where $S$ is the Hamilton principal function. 
If the Hamilton-Jacobi equation allows a separable solution, then it takes the form~:
\begin{equation}
\label{geodesics: Hamilton-Jacobi ansatz}
    S = \frac{1}{2} \chi \tau - E t + L \phi + S_\rho (\rho) + S_z (z).
\end{equation}
Inserting this Ansatz into (\ref{geodesics: Hamilton-Jacobi equation}), we find~:
\begin{equation}
    \left( \partial_\rho S_\rho \right)^2 - (1+j^2\rho^4)\chi + \frac{L^2 (1+ j^2\rho^4)^2}{\rho^2} + \left( \partial_z S_z \right)^2 - \left( E + 4jLz \right)^2 = 0,
\end{equation}
which leads to the equations~:
\begin{eqnarray}
    \rho^2 \left(\partial_\rho S_\rho \right)^2 = R(\rho), &\qquad& R(\rho) =  \rho^2 (k+F\chi)-L^2F^2; \nonumber \\
    \\
    \left( \partial_z S_z \right)^2 = \Xi(z), &\qquad& \Xi(z) = -k + \left( E+\omega L \right)^2, \nonumber
\end{eqnarray}
where $k$ is the separation constant, itself a constant of motion, which is akin to the Carter constant for the Kerr(-Newman) space-time \cite{carter1968global}.
The solution for $S$ then reads~:
\begin{equation}
    S = \frac{1}{2} \chi \tau - E t + L \phi + \int_{\rho} \frac{\sqrt{R}}{\rho}{\rm d}\rho + \int_{z} \sqrt{\Xi} {\rm d}z.
\end{equation}
The basic equations governing the motion can be deduced from Jacobi's principal function by the standard procedure of setting the partial derivative of $S$ with respect to the four constants of motions to zero. 
Alternatively, one can use the expressions for the generalised momenta, which read~:
\begin{equation}
    p_\mu = g_{\mu\nu}\dot{x}^\nu = \partial_\mu S \ .
\end{equation}

Using these, the motion in the swirling universe is completely determined by the system of first-order differential equations:
\begin{eqnarray}
\label{rho eq}   \frac{{\rm d} \rho}{{\rm d}\lambda}  &=& \xi_{\rho} \frac{\sqrt{R(\rho)}}{\rho}; \\
\label{z eq}     \frac{{\rm d} z}{{\rm d}\lambda} &=& \xi_{z} \sqrt{\Xi(z)}; \\
\label{t eq}      \frac{d t}{d\lambda}  &=& E+ 4jLz; \\
\label{phi eq}    \frac{d \phi}{d\lambda}  &=& \frac{L}{\rho^2}\left(1+ j^2\rho^4 \right)^2 -4jz \left(E+ 4jLz \right),
\end{eqnarray}
where we have defined the ``Mino time'' $\lambda$ by $F {\rm d}\lambda = {\rm d}\tau$ (a construction akin to that in \cite{mino2003perturbative}). 
The expressions $\xi_{\rho}=\pm 1$ and $\xi_{z}=\pm 1$ have been introduced to ensure the two possible choices of sign for the motion.
These can be chosen independently but must be kept coherently for the study of a given orbit.

\section{Complete set of solutions to the geodesic equations}
\label{solutions}

The general solution of the geodesic equations (\ref{rho eq}) - (\ref{phi eq}) is determined by the behavior of the polynomials $R(\rho)$ and $\Xi(z)$, which are fully characterized by the parameter $j$ and the four constants of motions: $E$, $L$, $\chi$ and $k$. 
The four geodesic equations can be analytically integrated using elementary functions, as well as the Weierstrass  $\wp-$, $\zeta-$, and $\sigma-$ functions, respectively. 
All solutions are given in terms of $\lambda$. 
In Appendix \ref{affine}, we give the explicit relation between $\lambda$ and the affine parameter $\tau$.

\subsection{$\rho$-motion}
\label{rho motion section}

The motion in the $\rho$-direction is described by the equation (\ref{rho eq}) and reads~:
\begin{equation}
\label{radial geodesic explicit}
   \rho^2 \left( \frac{\rm d \rho}{\rm d \lambda} \right)^2 = R(\rho) \ \ , \ \
   R(\rho)=\sum_{n=0}^{4} a_n \rho^{2n}  \ ,
\end{equation}
where
\begin{equation}
\label{rho equation series coefficients}
        a_0 = -L^2, \hspace{1cm} a_1=k+\chi, \hspace{1cm} a_2=-2j^2L^2, \hspace{1cm} a_3 = j^2\chi, \hspace{1cm} a_4=-j^4L^2 \ .
\end{equation}
This equation can be cast into a standard elliptical form by the following transformation~:  
\begin{equation}
\label{Weierstrass transformation X polynomial}
    \left( \frac{ {\rm d} q}{ {\rm d} \lambda} \right)^2 = Q(q) \ \ , \ \  \qquad Q(q) = 4 \sum_{n=0}^{4} a_n q^{n} = \sum_{n=0}^{4} \tilde{a}_n q^n \ ,
\end{equation}
where $Q(q)=4R(q)$ is a fourth-order polynomial in $q:=\rho^2$. 
The differential equation above has a mathematical structure similar to the equation describing the radial motion in the Reissner-Nordstr\"om and Kerr(-Newman) space-time, respectively, for both uncharged and charged particles \cite{hackmann2010geodesic,grunau2011geodesics,hackmann2013charged}.

Here, we give the basic steps and refer the reader to Appendix \ref{weierstrass_form} for more details.
First, we define a new variable $u$ via $q = \frac{1}{u} + q_0$, where $q_0$ is a root of $Q(q)$. 
The differential equation (\ref{Weierstrass transformation X polynomial}) then reads
\begin{equation}
\left( \frac{{\rm d}u}{{\rm d}\lambda} \right)^2 = \sum_{j=0}^{3} b_j u^{j} = P_3(u) \ ,
\end{equation}
where $P_3(u)$ is a third order polynomial in $u$. 
Applying a further coordinate transformation $u = \frac{1}{b_3} \left(4v-\frac{b_2}{3}  \right)$ we get
\begin{equation}
\label{geodesics: weierstrass form}
        \left( \frac{{\rm d}v}{{\rm d}\lambda} \right)^2 = 4 v -g_2 v - g_3 =P_W(v) \ .
\end{equation}
The right hand side of the equation is given such that the solution to (\ref{geodesics: weierstrass form}) can be given in terms of the Weierstrass $\wp$-function~:
\begin{equation}
    v(\lambda) = \wp\left( \lambda - \lambda_{in}^{(\rho)}; g_2, g_3 \right),
\end{equation}
where $\lambda_{in}^{(\rho)}$ depends exclusively on the initial conditions as follows~:
\begin{equation}
    \lambda_{in}^{(\rho)} = \lambda_0 + \xi_{\rho} \int_{v_0}^{\infty} \frac{\rm d v'}{{\rm \sqrt{P_W(v')}}}, \qquad
    v_0= \frac{1}{4} \left( \frac{b_3}{\rho_{in}^2 - q_0 } + \frac{b_2}{3} \right), 
\end{equation}
$\rho_{in} = \rho(\lambda_0)$ is the initial radial value for a given orbit, and the Weierstrass invariants are:
\begin{equation}
\label{Weierstrass invariantes}
        g_2 = -\frac{1}{4} \left( b_1 b_3 - \frac{b_2^2}{3} \right), \qquad g_3 = - \frac{1}{16} \left( b_0 b_3^2 + \frac{2 b_2^3}{27} - \frac{b_1 b_2 b_3}{3} \right).
\end{equation}
Therefore, the solution to the radial geodesic equation (\ref{rho eq}) is:
\begin{equation}
\label{rho equation: rho of lambda}
        \rho(\lambda) = \sqrt{\frac{b_3}{4 \wp \left( \lambda - \lambda_{in}^{(\rho)}; g_2, g_3 \right) - \frac{b_2}{3}} + q_0} \ .
\end{equation}

\subsection{$z$-motion }

The motion in the $z$-direction is described by (\ref{z eq}) and reads~:

\begin{equation}
\label{eq:zeq}
     \frac{d z}{d\lambda}  = \xi_z \sqrt{ -k + \left( E + 4jL z \right)^2 } \ .
\end{equation}
For $k > 0$ the $z$-motion is restricted to
\begin{equation*}
z \leq  z_{-}=\frac{-(\sqrt{k} \vert j\vert \vert L\vert + j EL)}{4 j^2  L^2} \ , \ \ {\rm and} \ \ z \geq z_{+}=\frac{\sqrt{k} \vert j\vert \vert L\vert - j EL}{4 j^2 L^2} \ , 
\end{equation*}
with the equal signs defining the turning points. The equation can be directly integrated:
\begin{equation}
    \int_{\lambda_0}^{\lambda} {\rm d}\lambda' = \frac{\xi_z}{4jL} \int_{\tilde{z_0}}^{\tilde{z}} \frac{{\rm d}\tilde{z}'}{\sqrt{-1+ \tilde{z}'} \sqrt{1+\tilde{z}'}} = \frac{\xi_z}{4jL} \left. \cosh^{-1}(\tilde{z}') \right|^{\tilde{z}}_{\tilde{z_0}},
\end{equation}
with $\tilde{z}=\frac{E+4jLz}{\sqrt{k}}$, hence:
\begin{equation}
    \lambda-\lambda_{in}^{(z)}=\frac{1}{4jL} \cosh^{-1} \left( \frac{E+4jLz}{\sqrt{k}} \right),
\end{equation}
 where $\lambda_{in}^{(z)}=\lambda_0 - \frac{\xi_z}{4j \text{L}} \cosh^{-1} \left( \frac{E+4j \text{L} z_{in}}{\sqrt{k}} \right)$ and $z_{in}=z(\lambda_0)$ is the initial value of $z$ for a given orbit.
Thus, one finds:
\begin{equation}
\label{z equation: z of lambda}
    z(\lambda) = \frac{1}{4jL} \left( \sqrt{k} \cosh \left( 4jL \left( \lambda-\lambda_{in}^{(z)} \right) \right) - E \right) \ .
\end{equation}

\subsection{$t$-motion}

The motion in $t$-direction is given by the equation (\ref{t eq}) and reads~:
\begin{equation}
    \frac{{\rm d} t}{{\rm d} \lambda} = E + 4jLz.
\end{equation}
Using (\ref{z equation: z of lambda}) this equation can be straightforwardly integrated, and one finds:
\begin{equation}
    t(\lambda) = \frac{\sqrt{k}}{4j\text{L}} \sinh \left( 4 j \text{L} \lambda \right) + t_{in},
\end{equation}
where $    t_{in} = t_0 - \frac{\sqrt{k}}{4j\text{L}} \sinh \left( 4 j \text{L} \left( \lambda_0 - \lambda_{in}^{(z)} \right) \right)$ with $t_{0}=t(\lambda_0)$ the initial value for the time coordinate.

\subsection{$\phi$-motion}

The motion in $\phi$-direction is given by the equation (\ref{phi eq}) and reads~:
\begin{equation}
    \frac{\rm d \phi}{d\lambda}  = \frac{L}{\rho^2}\left(1+ j^2\rho^4 \right)^2 -4jz \left(E+ 4jLz \right),
\end{equation}
and therefore can be integrated considering the two contributions:
\begin{equation}
\label{phi motion - separated integrals}
    \int_{\phi_{in}}^{\phi} {\rm d}\phi' = \phi(\lambda) - \phi_{in} = \mathcal{I}_{\rho} - \mathcal{I}_z,
\end{equation}
where $\phi(\lambda_0)=\phi_{in}$ is the initial value of the $\phi$ component, and:
\begin{equation}
\label{phi contributions}
    \mathcal{I}_\rho = \text{L} \int_{\lambda_0}^{\lambda} \frac{(1+j^2\rho^4)^2}{\rho^2} {\rm d}\lambda;
    \qquad
    \mathcal{I}_z = 4j \int_{\lambda_0}^{\lambda} z(E+4jLz) {\rm d}\lambda.   
\end{equation}
These two contributions can be directly integrated using the solutions (\ref{rho equation: rho of lambda}) and (\ref{z equation: z of lambda}). 
The first is then cast into an elliptic integral of the third kind, see Appendix \ref{integration of elliptic integrals of the third kind} for more details.
The second one can be integrated directly.

First, note that the integral $\mathcal{I}_{\rho}$ can be rewritten by performing the same set of transformations as described in Section \ref{rho motion section}, thus it becomes:
\begin{equation*}
\label{elliptic integral of third kind}
    \mathcal{I}_\rho = \int_{v_0}^{v}  f(v') \frac{dv'}{\sqrt{P_W(v')}} ,
\end{equation*}
where the function $f(v')$ written in terms of the partial fraction decomposition is:
\begin{eqnarray}
    f(v) &=& K_0  +\frac{K_1}{v-\alpha} + \frac{K_2}{(v-\alpha)^2} + \frac{K_3}{(v-\alpha)^3} + \frac{C_1}{v-\beta},
\end{eqnarray}
where $\alpha = \frac{b_2}{12}$ and $\beta = \frac{b_2}{12} - \frac{b_3}{4 q_0}$ are the roots of $D(v) =(b_2-1 2v)^3 (-3b_3 + (b_2-12v)q_0) $, and the coefficients are:
\begin{eqnarray*}
    && K_0= \frac{\text{L} \left( 1+j^2 q_0^2 \right)^2}{q_0}, \qquad K_1 = \frac{3 b_3 \text{L} \left( 2 + 3 j^2 q_0^2  \right)j^2}{12}, \qquad C_1 = \frac{3 b_3 \text{L}}{12 q_0^2} , \\ 
    && K_2 = \frac{27 j^4 b_3^2 q_0 \text{L}}{144}, \qquad K_3 = \frac{27 b_3^2 j^4 \text{L}}{1728}.
\end{eqnarray*}

From Section \ref{rho motion section} we know that the solution of ${\rm d}\lambda = \frac{{\rm d}v}{\sqrt{P_W}}$ is given by $v(\lambda) = \wp \left( \lambda - \lambda_{in}^{(\rho)} \right)$. Thus $f(\lambda) = f(\wp(\lambda))$ is an elliptic function with the same half-periods of $\wp(\lambda)$ and therefore is an elliptic integral, whose solution is:
\begin{eqnarray}
\label{phi rho component}
    \mathcal{I}_\rho &=& \gamma_0 \left( \lambda - \lambda_0 \right) + 
    %Adding wp
                \gamma_1 \left( \wp(\lambda - \lambda_{in}^{(\rho)} + y_\alpha) - \wp(\lambda - \lambda_{in}^{(\rho)} - y_\alpha) - \wp(\lambda_0 - \lambda_{in}^{(\rho)} + y_\alpha) + \wp(\lambda_0 - \lambda_{in}^{(\rho)} - y_\alpha) \right)  
                \nonumber \\
    %Adding zetas
                & & + \gamma_2 \left[ \zeta\left( \lambda - \lambda_{in}^{(\rho)} - y_\alpha\right) + \zeta \left( \lambda - \lambda_{in}^{(\rho)} + y_\alpha \right) -\zeta\left( \lambda_0 - \lambda_{in}^{(\rho)} - y_\alpha\right) - \zeta \left( \lambda_0 - \lambda_{in}^{(\rho)} + y_\alpha \right) \right] 
                \nonumber \\
    %Adding ln y beta
              & &  +  \gamma_3 \left[  \ln \left( \frac{\sigma(\lambda - \lambda_{in}^{(\rho)}- y_\beta)}{\sigma(\lambda - \lambda_{in}^{(\rho)}+ y_\beta)}\right) - 
                \ln \left( \frac{\sigma(\lambda_0 - \lambda_{in}^{(\rho)}- y_\beta)}{\sigma(\lambda_0 - \lambda_{in}^{(\rho)}+ y_\beta)} \right)\right] \nonumber \\
    %Adding ln y alpha
              & & +  \gamma_4  
                    \left[  \ln \left( \frac{\sigma(\lambda - \lambda_{in}^{(\rho)}- y_\alpha)}{\sigma(\lambda - \lambda_{in}^{(\rho)}+ y_\alpha)}\right) - 
                    \ln \left( \frac{\sigma(\lambda_0 - \lambda_{in}^{(\rho)}- y_\alpha)}{\sigma(\lambda_0 - \lambda_{in}^{(\rho)}+ y_\alpha)} \right)\right] \ ,
\end{eqnarray}
where the values of the constants are given by:
\begin{eqnarray*}
    \gamma_0 &=& \left[  K_0 + \frac{2 \zeta(y_\alpha) K_1}{\wp'(y_\alpha)} + \frac{2 \zeta(y_\beta) C_1}{\wp'(y_\beta)} - \frac{K_3}{\wp'(y_\alpha)^2} \left( 1 + \frac{12 \wp(y_\alpha) \zeta(y_\alpha)}{\wp'(y_\alpha)} \right) \right.    
    \nonumber \\
    & & \left. + \frac{1}{\wp'(y_\alpha)^2}\left(\wp(y_\alpha) + \frac{\wp''(y_\alpha)\zeta(y_\alpha)}{\wp'(y_\alpha)} \right) \left( \frac{3K_3 \wp''(y_\alpha)}{\wp'(y_\alpha)^2} - 2K_2 \right)
    \right] \ ,
    \\
    \gamma_1 &=& \frac{K_3}{2 \wp'(y_\alpha)^3} \ , \\
    \gamma_2 &=& \frac{1}{\wp'(y_\alpha)^2} \left(-K_2 + \frac{3 K_3 \wp''(y_\alpha)}{2\wp'(y_\alpha)} \right) \ , \\
    \gamma_3 &=& \frac{C_1}{\wp'(y_\beta)} \ , \\
    \gamma_4 &=& \frac{K_1}{\wp'(y_\alpha)} + 
            \frac{K_2 \wp''(y_\alpha)}{\wp'(y_\alpha)^3} - 
            \frac{3 K_3}{ \wp'(y_\alpha)^3} \left( 2 \wp(y_\alpha) - \frac{\wp''(y_\alpha)^2}{2 \wp'(y_\alpha)^2} \right) \ , 
\end{eqnarray*}
and $y_\alpha$ and $y_\beta$ are values of the inverse of the Weierstrass $\wp$-function. 
Hence $\wp(y_\alpha) = \alpha$ and $\wp(y_\beta) = \beta$. $\zeta(y)$ and $\sigma(y)$ are, respectively, the Weierstrass $\zeta$- and $\sigma$-function. 

Note that special attention is required when evaluating the logarithm in (\ref{phi rho component})  in order to produce a continuous implementation of $\phi(\lambda)$ ensuring that we use the strategy discussed in \cite{kerrweierstrass}. 

Now, we consider the integral $\mathcal{I}_z$, which can be directly integrated using (\ref{z equation: z of lambda}). One finds:
\begin{eqnarray}
\label{phi z component}
    \mathcal{I}_z = \frac{1}{16 j L^2} \left[ 8jLk (\lambda-\lambda_{0}) + k\left(\sinh(8jL\lambda) - \sinh(8jL\lambda_0)\right) - 4E\sqrt{k}\left(\sinh(4jL\lambda) - \sinh(4jL\lambda_0)\right) \right] \ .
\end{eqnarray}
Hence, the solution of the geodesic equation in $\phi$-direction is fully described by (\ref{phi motion - separated integrals}) together with (\ref{phi rho component}) and (\ref{phi z component}).

\section{Examples of orbits}
\label{Orbits}
Using the complete set of solutions to the geodesic equations given in Section \ref{solutions}, we can now present examples of orbits in the swirling universe space-time.

The motion of particles is characterized by the constants of motion, $E$, $L$, and $k$, together with the normalization condition, $\chi$. 
Inspection of the equations for $\rho$ (\ref{rho eq}) and $z$ (\ref{z eq}) gives us information on the possible choices of these constants. (\ref{rho eq}) leads to the inequality
\begin{equation}
    k \ge \frac{L^2 F^2}{\rho^2} - \chi F ,
    \label{rho in}
\end{equation}
and since $-\chi F \geq 0 $ everywhere, this gives a lower bound on $k$. Besides it corroborates that $k$ must be positive. Eq.~(\ref{z eq}), on the other hand, leads to the inequality
\begin{equation}
    k \le (E + 4 j z L)^2 ,
    \label{z in}
\end{equation}
which puts an upper bound on $k$.
If the angular momentum of a massive particle vanishes, $L=0$, the combined bounds reduce to $1+j^2\rho^4 \le k \le E^2$.
This corresponds to a turning point in $\rho$ and no restriction in $z$.
In particular, the equatorial plane can be traversed in this oscillatory motion.
For finite angular momentum $L$, on the other hand, there will also be an inner turning point for the $\rho$-motion that depends on $k$, since the inequality (\ref{rho in}) contains a $1/\rho^2$ term. 
Then, the only motion possible is that between these two turning points.
Moreover, the inequality (\ref{z in}) can give two turning points in the $z$-motion such that only motion outside these two turning points is possible.
A typical orbit will thus oscillate in $\rho$-direction between the two turning points and escape to infinity in $z$-direction. 
A \textit{Wolfram Mathematica} notebook implemented to plot the orbits described in the above section is available at \cite{github notebook}.

We find that there are only two possibilities for the motion in $z$ direction depending on the sign of the initial velocity: (a) considering a particle starting at $z_0 \geq z_+$ with an initial velocity $\dot{z}>0$ the particle escapes directly to $+ \infty$, (b) if it has an initial velocity $\dot{z}<0$ it moves until it reaches the turning point at $z_+$ and then escapes to $+\infty$. The description is completely analogous for a particle that starts at $z_0 \leq z_-$.

The turning points of $R(\rho)$ define the regions where motion can exist in $\rho$-direction. Since this is a bi-quadratic eighth-order polynomial, all zeros have, at least, multiplicity two. 
Thus, there is no loss of generality by studying the zeros of the equivalent fourth-order polynomial. 
This can have two or four real zeros and only the positive real roots have physical relevance. 
The full classification in that direction is discussed below for the motion of massless and massive particles, respectively.

\subsection{Massless particles}

For massless particles we have $\chi=0$, thus the motion in the $\rho$-direction is described by~:
\begin{equation}
    \left( \frac{{\rm d} q}{{\rm d}\lambda} \right)^2 = -j^4 L^2 q^4 - 2j^2 L^2 q^2 +k q - L^2 =Q(q).
\end{equation}

\begin{figure}[h!]
    \centering
    \includegraphics[scale=.6]{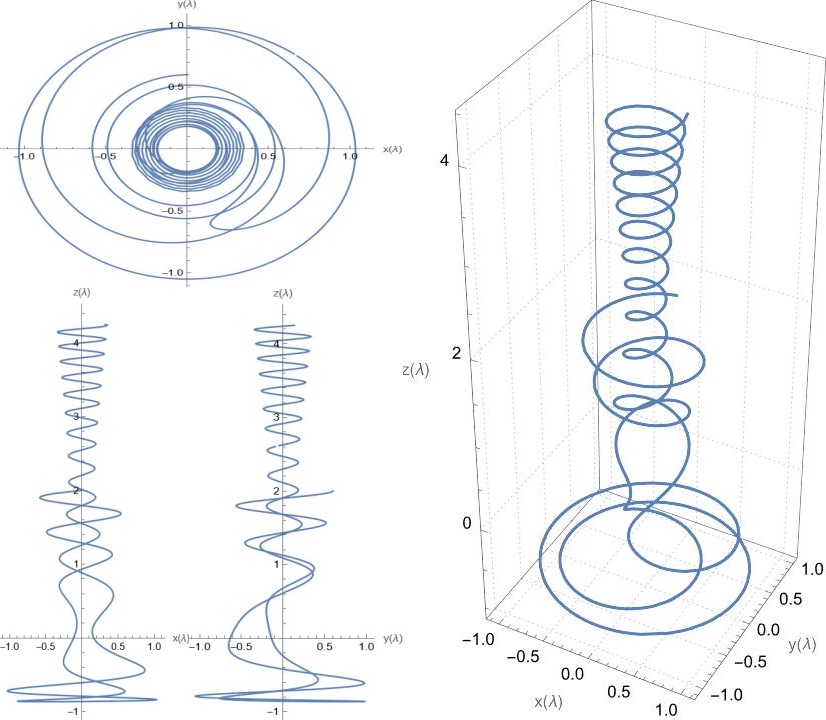}
    \includegraphics[scale=.6]{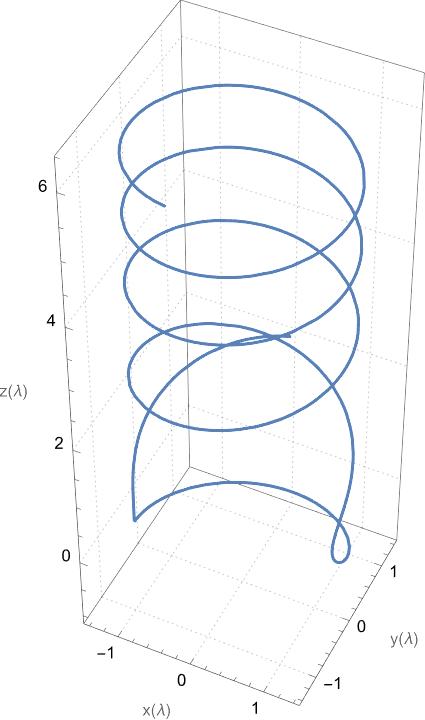}
    \caption{We show examples of orbits for massless particles. 
    The left and middle figures are for $j=2$, $L=0.4$ and $k=5$, and $E=5$ and hence the radial motion has two turning points (see the discussion in the text). 
    The left figure shows the projection of the orbit onto the $x$-$y$-, $x$-$z$- and the $y$-$z$-plane, respectively, the middle figure shows the motion in $3$ dimensions. The right figure is for $j=\sqrt{\frac{27}{256}}$, $L=1$ and $k=1$, and $E=2$, such that the only possible motion is when choosing the initial condition as $\rho_0 = \frac{4}{3}$. 
    Note that both orbits have $ \dot{z}(0)<0 $, and hence the particle starting at $z(0) = 2$ moves downwards towards the turning point $z_+$ and then escapes to infinity.
    }
    \label{Massless plot}
\end{figure}

The regions where orbits are allowed to exist will strongly depend on the zeros of the above polynomial. 
Orbits can exist only between two real zeros, where the polynomial $Q(q)$ has positive values. The discriminant of the above polynomial is:
\begin{equation}
    \Delta = j^6 \xi k^2 \left( 256 \xi - 27 k^2 \right) \ .
\end{equation}
There are always two distinct real roots for $\xi =j^2 L^4 < \frac{27}{256}k^2$. Orbits exist when these roots are positive. 
Then the radial coordinate oscillates between these two values. An orbit of this type is shown in Fig.~\ref{Massless plot}. 
For $\xi = \frac{27}{256}k^2$, the function $Q(q)$ has multiple roots; however, since $Q''(q)$ is negative, it will turn back again to negative values. 
Therefore the only possible motion is an orbit with constant radius: $\rho(\lambda) = \frac{4L}{3\sqrt{k}}$. 
Such an orbit is shown in Fig.~\ref{Massless plot}. 
Moreover, for the cases $\xi > \frac{27}{256}k^2 $ and $k=0$, respectively, there are no real turning points of the polynomial $Q(q)$, which is negative for all values of $q>0$, and thus no orbits of massless particles are allowed.

\subsection{Massive particles}

For massive orbits, one has $\chi=-1$, thus the motion in $\rho$-direction is described by the equation~:
\begin{equation}
    \left(  \frac{{\rm d} q}{{\rm d}\lambda} \right)^2 = -j^4 L^2 q^4 -j^2 q^3 - 2j^2 L^2 q^2 +(k-1) q - L^2 =Q(q).
\end{equation}
The zeros of the polynomial $Q(q)$ define the regions where $Q(q)\geq 0$ and hence enclose the regions where orbits are allowed to exist.   

The polynomial $Q(q)$ can either have four real distinct roots, two real distinct and two complex conjugate roots, two complex conjugate pairs of roots, or multiple (real and complex) roots. 
Physical orbits only exist between two real positive roots. 
The existence of zeros can be studied by making use of the discriminant:
\begin{equation}
    \Delta_{\xi}(k) = j^6 \left(-27 \xi k^4 + \left(72 \xi + 4\right) k^3 + 4 \left(64 \xi^2 -14 \xi^4-3\right) k^2 - 4 \left(8 \xi^4-3\right) k + 16 \xi -4 \right) \ ,
\end{equation}
which can be studied as a fourth order polynomial in $k$.
Thus, there are different combinations of $k$ and $\xi=j^2 \text{L}^4 > 0$ for which the above discriminant can be positive, negative, or zero. 

\begin{figure}
    \centering
    \includegraphics[scale=0.6]{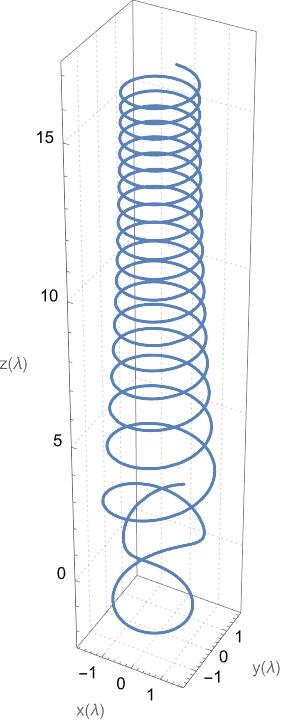}
    \includegraphics[scale=0.6]{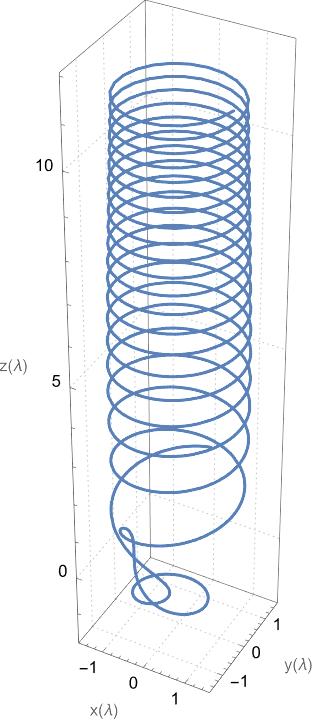}
    \includegraphics[scale=.6]{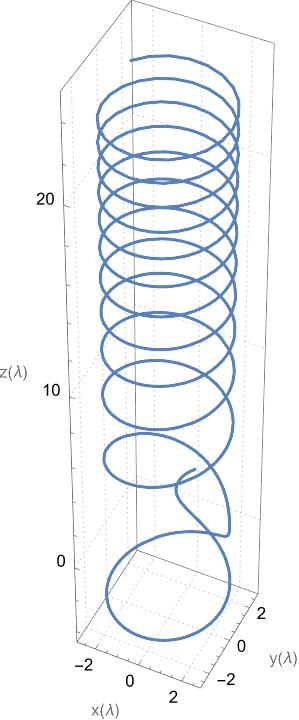}
    \caption{Examples of orbits for massive particles. 
    The figure on the left shows an orbit from \textbf{Region 1} with $j=\sqrt{1/20}$, $L=1$, $k = 2.2$ and $E=3$. 
    The figure in the middle shows an orbit from \textbf{Region 2} with $j=0.5$, $L=1.2$, $k = 5$ and $E = 5$. 
    For these two cases the radial motion is restricted to take place between the turning points. 
    The figure on the right shows an orbit from \textbf{Region 3} with $j=1/16$, $L=2$, $k = 2$ and $E = 3$ for which we have to choose $\rho(0)=4\sqrt{\sqrt{2}-1}$ such that $\rho$ stays constant throughout the motion.}
    \label{massive plot}
\end{figure}

Therefore, we distinguish the regions of allowed orbits as follows~:

\begin{itemize}
    \item \textbf{Region 1 - $\Delta_{\xi}(k) > 0$}: Four real distinct roots of $Q(q)$ can exist. 
    There are two different possibilities: (a) $k<1$, then $ 0 < \xi <\frac{3}{16}$, but $\xi \neq \frac{1}{16}$ and $k_1 < k < k_2 $, and (b) $k>1$ and $\xi<\frac{3}{16(k+1)}$ but $\xi \neq \frac{1}{16}$ and $k_1 < k < k_2 $, notice that $k=2$ is excluded for this region. 
    An example of such an orbit is shown in Fig. \ref{massive plot} (left).
    
    \item \textbf{Region 2 - $\Delta_{\xi}(k) < 0$}: Two distinct real roots of $Q(q)$. 
    There are three possibilities: (a) $\xi > \frac{1}{4} $ and $k>k_2$, (b) $\xi \leq \frac{1}{4}$ and $0<k<k_1$ or $k > k_2$, and (c) $\xi = \frac{1}{16}$ and $k>2$. 
     An example of such an orbit is shown in Fig. \ref{massive plot}  (middle).

    \item \textbf{Region 3 - $\Delta_{\xi}(k) = 0 $}: $Q(q)$ has multiple roots. 
    This region is accessible once $k$ is chosen to be a root of $\Delta_{\xi}(k)$ for a given value of $\xi$. 
    However, since $Q(q)$ only reaches zero and then turns again to be negative, the only possible orbit is $\rho(\lambda)=\rho(0)$, where $\rho_0$ is the root of $R(\rho)$. 
    A special choice satisfying this is $\xi=\frac{1}{16}$ and $k=2$, and then the motion is allowed for $\rho(\lambda) = 2\vert L\vert\sqrt{\sqrt{2}-1}$. 
    An example of such an orbit is shown in Fig.~\ref{massive plot} (right).
\end{itemize}

\section{Geodesic motion in a space-time describing a black hole in a swirling universe}
\label{BHswirling}

\begin{figure}[h]
    \centering
    \includegraphics[scale=0.6]{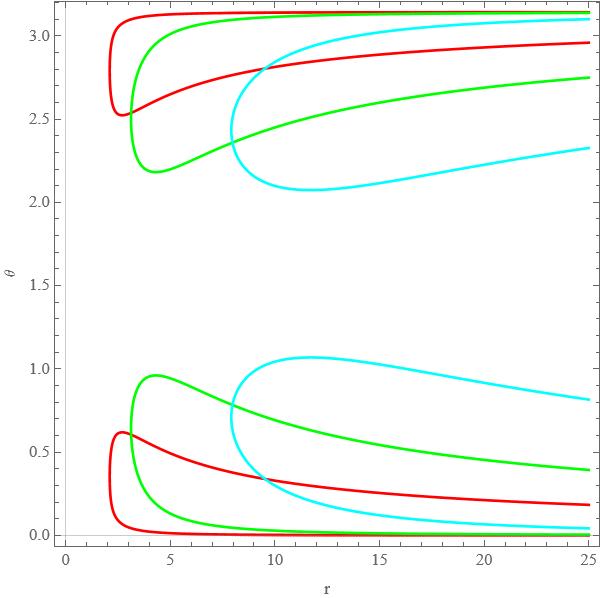}
    \includegraphics[scale=0.7]{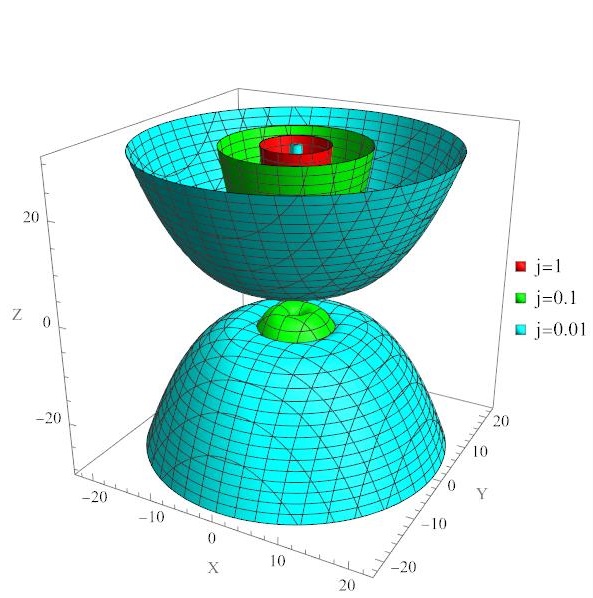}
    \caption{We show the ergoregions for the space-time describing a Schwarzschild black hole of mass $M=1$ immersed in a swirling universe for three different values of $j$. 
    On the left, we give the projection of the ergoregions onto the $r$-$\theta$-plane, while the right figure shows the ergoregions rotated around the symmetry axis.
    }
    \label{figure black hole ergoregion}
\end{figure}

Application of the Ehlers' transformation using a black hole as a seed leads to a solution describing a black hole in a swirling universe \cite{astorino2022black}. 
This works akin to applying the Harrison transformation to a black hole seed leading to a black hole immersed in a Melvin magnetic universe. 
Using the Ehlers' transformation with a Schwarzschild black hole seed, the following metric was presented in spherical coordinates $(t,r,\theta,\varphi)$ \cite{astorino2022black}~:
\begin{equation}
\label{bh metric tensor}
    {\rm d}s^2 = F(r,\theta) \left( - N(r) {\rm d}t^2 + \frac{{\rm d}r^2}{N(r)} + r^2 {\rm d}\theta^2 \right) + \frac{r^2 \sin^2\theta}{F(r,\theta)} \left( {\rm d}\varphi + \omega(r,\theta) {\rm d}t \right)^2 \ ,
\end{equation}
where $F(r,\theta)=1+j^2 r^4 \sin^4\theta$, $N(r)=1-\frac{2M}{r}$ and $\omega(r,\theta)=4j (r-2M)\cos\theta + \omega_0$. For $M=0$, this space-time reduces to the space-time (\ref{eq:background}). 
For $M\neq 0$, it has an event horizon at $r=2M$, which for $j\neq 0$ is prolate-shaped rather than perfectly spherically symmetric as in the Schwarzschild case. 
Besides, similarly to the background, this space-time also has an ergoregion defined by
\begin{equation} 
-F^2(r,\theta) N(r) + r^2\sin^2\theta \omega(r,\theta)^2 = 0  \ .
\end{equation}
Note that the event horizon $r=2M$ fulfills this relation, but is not an ergosurface.
We hence require for the ergoregion that $r > 2M$.
In Fig.\ref{figure black hole ergoregion} we show the ergoregions for this space-time for $M=1$ and three different values of $j$.

Now, trying to solve the geodesic equations, we note that unlike 
(\ref{eq:background}) the space-time (\ref{bh metric tensor}) is of Petrov type I, and therefore it is not expected to have additional constants of motion.
Hence, we will not be able to separate the geodesic equations fully. 
The geodesic Lagrangian reads~:
\begin{equation}
\label{lagragian bh space}
    2 \mathcal{L} = F(r,\theta) \left( - N(r) \dot{t}^2 + \frac{\dot{r}^2}{N(r)} + r^2 \dot{\theta}^2 \right) + \frac{r^2 \sin^2\theta}{F(r,\theta)} \left( \dot{\varphi} + \omega(r,\theta) \dot{t} \right)^2 = \chi,
\end{equation}
where we have set $\omega_0=0$, since it does not influence the geodesic motion.
There are two constants of motion given by the cyclic variables, which are the particle's total energy $E$, and angular momentum $L$, Eq.~(\ref{geodesics: conserved quantities cyclic variables}). 
Solving for $\dot{t}$ and $\dot{\phi}$ we find:
\begin{eqnarray}
    \dot{t} &=& \frac{r\left( E + 4jL(r-2M)\cos\theta \right)}{(r-2M)(1+j^2r^4\sin^4\theta)}, \nonumber \\
     \dot{\varphi} &=& \frac{L - j r^3 \sin^2\theta \left( 4\cos\theta \left( E-8jML\cos\theta \right)-Lj^3r^5\sin^6\theta + 2jLr(9\cos^2\theta-1)\right)}{r^2\sin^2\theta (1+j^2r^4\sin^4\theta)}  \  .
\end{eqnarray}

A description of the motion in this space-time requires a full numerical integration of the coupled system of geodesic equations. 
Here we present some preliminary results with a full analysis being presented elsewhere \cite{capobianco_hartmann_kunz}.

\begin{figure}[p]
\label{figure: bh space-time geodesic}
\begin{minipage}{8cm}
    \centering
    \subfigure[Orbit in the $x$-$y$-plane]{\includegraphics[scale=.6]{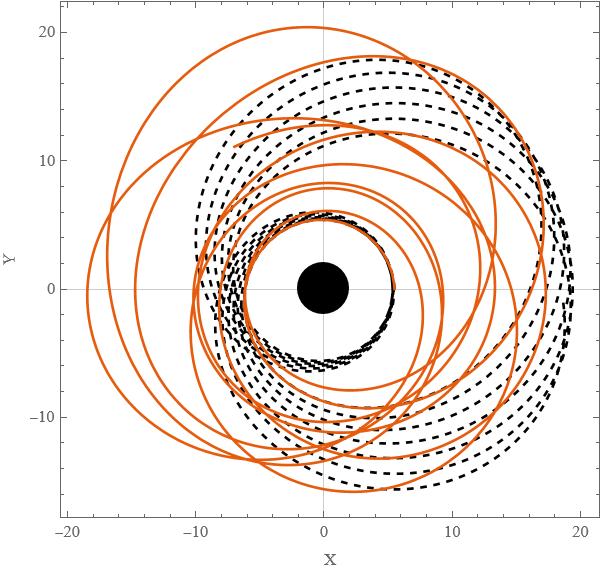}\label{bh massive projection}}
    \subfigure[Orbit 1 - Bounded orbit]{\includegraphics[scale=.5]{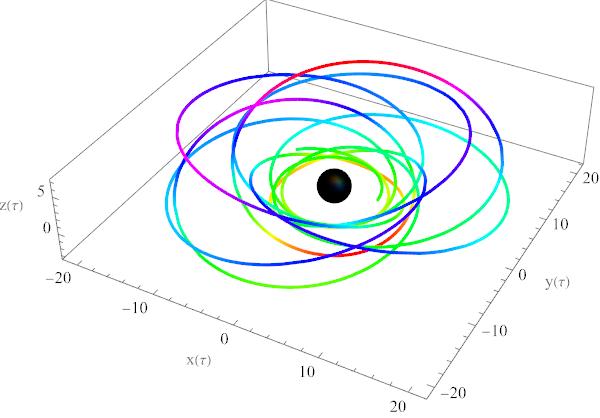}\label{bh orbit 1}}
    \includegraphics[scale=.9]{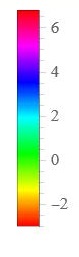}
\end{minipage}
\begin{minipage}{8cm}
    \centering
    \subfigure[Orbit 2 - Escape Orbit]{\includegraphics[scale=.7]{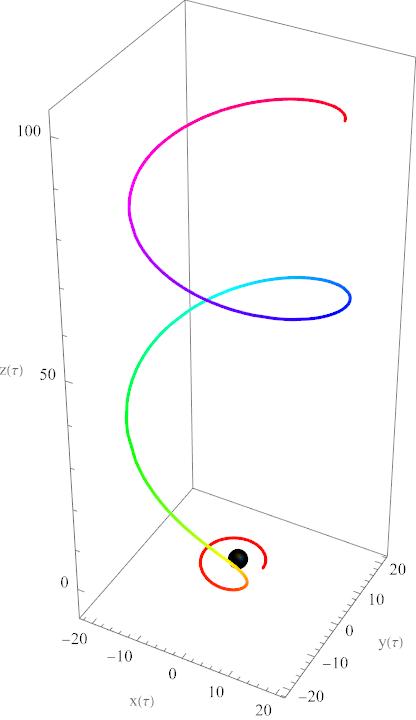}\label{bh orbit 2}}
    \includegraphics[scale=.9]{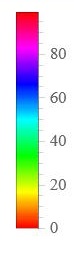}
\end{minipage}
\caption{Two time-like orbits sharing the same constants of motion $\text{E}=\sqrt{0.93}$ and $\text{L}=\sqrt{\frac{1}{0.072}}$. 
The initial conditions were chosen such that the positions are the same. 
Moreover, the initial polar velocity and the initial radial velocity are defined to satisfy the normalization condition for each case. 
The top left figure (a) shows a bound orbit for $M=1$, $j=4\times 10^{-5}$ in the $x$-$y$-plane (solid orange). 
For comparison the orbit in the Schwarzschild space-time ($M=1$, $j=0$) is also shown (black dashed). 
Figure (b) shows the orbit of figure (a) in $3$ dimensions.
Figure (c) shows an escape orbit with $M=1$ and $j=4\times10^{-4}$.
The colours in figure (b) and (c), respectively, indicate the location above or below the equatorial plane ($z=0$), while the surface of the black sphere indicates the horizon of the black hole. 
        }
\end{figure}

Interestingly, we find that even a small deviation of the swirling parameter $j$ from zero can change the qualitative features of the orbits significantly as compared to those in the Schwarzschild space-time. 
In Fig.~\ref{bh massive projection} we show a bound orbit of a massive particle for $M=1$ and $j=4\times 10^{-5}$ in comparison to the bound orbit obtained for $M=1$, $j=0$, i.e.~in the Schwarzschild space-time. 
The initial conditions were chosen such that the orbits are starting on the equatorial plane with $\dot{\theta}(0)=0$, and then $\dot{r}(0)$ is found by satisfying the normalization condition for each case. 
Note that while the Schwarzschild orbit (dashed black) is in the equatorial plane and has a  perihelion shift, this is very different for $j\neq 0$ (orange solid line). 
The orbit is non-planar (see  Fig.~\ref{bh orbit 1}) and shows no regular behaviour.
In Fig.~\ref{bh orbit 2} we show an orbit for $M=1$ but for $j=4\times 10^{-4}$. 
In contrast to the cases with $j=0$ and $j=4\times 10^{-5}$, respectively, this orbit is no longer bounded and escapes to infinity in $z$-direction.

\section{Conclusions}

We have considered geodesics in a swirling universe with and without a black hole immersed into it, which was obtained recently via an Ehlers' transformation by Astorino et al.~\cite{astorino2022black}.
We have focused on the case of the pure swirling universe, i.e., without the black hole, since in this case the geodesic equations can be solved analytically. 
In addition to the two cyclic variables present in any axially symmetric geometry, and the normalization condition, a fourth constant of motion has been obtained by making use of the Hamilton-Jacobi formulation. In this formalism, the four geodesic equations can be completely uncoupled and solved using elementary and elliptic functions.

The geodesic equations themselves allow already for a qualitative analysis of the types of motion possible, showing that the motion in $\rho$-direction is bounded, whereas the motion in $z$-direction is unbounded, unless the angular momentum of the particle vanishes.
We have presented a number of examples of orbits for massless and massive particles, illustrating their spiraling motion.
The rotational direction of this spiraling motion is seen to change, when the ergoregions are approached that are featured by the swirling universe \cite{gibbons2013ergoregions}.

When immersing a black hole into the swirling universe the space-time is no longer of Petrov type D. In that case, the geodesic equations can no longer be decoupled and solved by well-known analytical techniques.
Therefore we have obtained sample solutions numerically, showing that already rather small values of the swirling parameter $j$ will produce substantial changes with respect to the Schwarzschild orbits for vanishing $j$.

Next, we will study the geodesics in the swirling universe with a black hole immersed inside in full detail and analyze the possible types of motion for massless and massive particles.
The non-separability of the geodesic equations suggests that chaotic motions is present in this space-time. Thus, a full description of the motion will require a qualitative and quantitative classification of the emergence of chaos in this system as well.
Subsequently, the space-time describing a Kerr black hole immersed in a swirling universe \cite{astorino2022black} is waiting for analysis.

\section*{Acknowledgements}

R.C.~would like to thank Marco Astorino, Riccardo Martelli, and Adriano Vigan\`o for discussions.
He is grateful to CAPES for financial support under Grant No: 88887.371717/2019-00, and would like to thank the University of Oldenburg for hospitality. 
J.K. gratefully acknowledges support by DFG project Ku612/18-1.

\appendix

\section{Affine parameter}
\label{affine}

The relation between the Mino time $(\lambda)$ and the affine parameter $\tau$ reads~:
\begin{equation}
    \frac{{\rm d}\tau}{{\rm d}\lambda} = 1 + j^2 \rho(\lambda)^4  \ .
\end{equation}
Inserting (\ref{rho equation: rho of lambda}) we get
\begin{equation}
    \tau - \tau_0 = \int_{\lambda_0}^{\lambda} \left[  \epsilon_0 + \frac{\epsilon_1}{\wp \left( \lambda' - \lambda_{in}^{(\rho)} \right)  - \frac{b_2}{12} } + \frac{\epsilon_2}{\left( \wp \left( \lambda' - \lambda_{in}^{(\rho)} \right)  - \frac{b_2}{12} \right)^2 } \right] {\rm d} \lambda'
\end{equation}
with
\begin{eqnarray}
    \epsilon_0 = 1+j^2 q_0^2, \qquad \epsilon_1 = \frac{j^2 q_0 c_3}{2}, \qquad \epsilon_2 = \frac{j^2 c_3^2}{16}.
\end{eqnarray}
Thus
\begin{equation}
    \tau(\lambda) = \tau_0 + \epsilon_0 \left( \lambda - \lambda_0 \right) + \epsilon_1 \mathcal{I}_1(\lambda, y_\alpha) + \epsilon_2 \mathcal{I}_2 (\lambda, y_\alpha) .
\end{equation}

\section{Transformation to Weierstrass form}
\label{weierstrass_form}

The equation (\ref{Weierstrass transformation X polynomial}) allows for a solution in terms of the Weierstrass $\wp$-function. Here, we give all details of the necessary transformations. We start by reducing the polynomial 
$Q(q)$ from fourth to third order by the transformation
\begin{equation}
        q-q_0=\frac{1}{u} \ \ \ \Rightarrow \ \ \ {\rm d}q = -\frac{{\rm d} u}{u^2} \ ,
\end{equation}
where $q_0$ is a root of $Q(q)$. Thus, the differential equation becomes:
\begin{equation}
        \left( \frac{{\rm d} u}{{\rm d} \lambda} \right)^2 = P_3(u); \qquad P_3=\sum_{j=0}^{3} b_j u^j \ ,
\end{equation}
with the coefficients:
\begin{equation}
    b_0 = \tilde{a_4}, 
    \qquad b_1 = \tilde{a}_3 + 4 \tilde{a}_4 q_0,
    \qquad b_2 = \tilde{a}_2 + 3 \tilde{a}_3 q_0 + 6 \tilde{a}_4 q_0^2, 
    \qquad b_3 = \tilde{a}_1 + 2 \tilde{a}_2 q_0 + 3 \tilde{a}_3 q_0^2 + 4 \tilde{a}_4 q_0^3 \ .
\end{equation}

A general third-order polynomial can be cast into Weierstrass form by the transformation
\begin{equation}
       u = \frac{1}{b_3}\left( 4 v - \frac{b_2}{3} \right) \ \ \ \Rightarrow \ \ \
            \qquad { {\rm d} u} = \frac{4}{b_3}{{\rm d} v} \ ,
\end{equation}
thus giving~:
\begin{equation}
\label{eq:rho weierstrass form}
       \left( \frac{ {\rm d} v}{ {\rm d} \lambda} \right)^2 = 4 v^3 -g_2 v - g_3:= P_W(v),
\end{equation}
with
\begin{equation}
    g_2 = -\frac{1}{4} \left( b_1 b_3 - \frac{b_2^2}{3} \right), \qquad g_3 = - \frac{1}{16} \left( b_0 b_3^2 + \frac{2 b_2^3}{27} - \frac{b_1 b_2 b_3}{3} \right) \ .
\end{equation}
Hence, in addition to an initial value $v(\lambda_0)=v_0$ the solution is fully determined by:
\begin{equation}
         v(\lambda) = \wp\left( \lambda - \lambda_{in}, g_2, g_3 \right), \qquad \lambda_{in} = \lambda_0 + \int_{v_0}^{\infty} \frac{{\rm d} v'} {{\sqrt{P_W(v')}}} \ .
\end{equation}

\section{Integration of elliptic integrals of the third kind}
\label{integration of elliptic integrals of the third kind}

Here we give the formulae for the evaluation of integrals of the type: $\mathcal{I}_{n} = \int_{v_0}^{v} \frac{1}{\left( \wp(v')-\gamma \right)^{n}}$, with $n=1$,$2$ or $3$. Note that $\gamma = \wp(y_\gamma)$ is a single pole of the above expression. A table with these and other relations can be found in \cite{kerrweierstrass}.

The starting point is to consider the expansion of the denominator as:
\begin{equation}
\label{inverse wp expansion}
    \frac{\wp'(y)}{\wp(v)-\wp(y)} = \zeta\left( v - y \right) - \zeta\left( v + y \right) + 2 \zeta \left( y \right),
\end{equation}
which can then be directly integrated using $\ln \sigma(x) = \int \zeta(x) {\rm d}x$ to get:
\begin{equation}
\label{int I1}
        \mathcal{I}_{1}\left( v,y \right) = \int \frac{{\rm d}v}{\wp(v)-\wp(y)} = \frac{1}{\wp'(y)} \left[ 2 \zeta(y) v + \ln \frac{\sigma \left( v-y \right)}{\sigma \left( v+y \right)} \right].
\end{equation}  

Considering (\ref{inverse wp expansion}), taking the derivative with respect to $y$, and using $\frac{{\rm d}\zeta(y)}{{\rm d}y} =- \wp(y)$ one gets:
\begin{equation}
    \frac{1}{\left(\wp(v)-\wp(y)\right)^2} = \frac{1}{\wp'(y)^2} \left[ \wp(v-y) + \wp(v+y) +2\wp(y) - \frac{\wp''(y)}{\wp(v)-\wp(y)} \right] ,
\end{equation}
which can be directly integrated leading to:
\begin{equation}
\label{int I2}
        \mathcal{I}_{2} \left( v,y \right)= \int \frac{{\rm d}v}{\left( \wp(v)-\wp(y) \right)^2} =-\frac{\wp''(y)}{\wp'(y)^2} \mathcal{I}_1 - \frac{1}{\wp'(y)^2} \left( \zeta \left( v+y \right) + \zeta \left( v-y \right) + 2\wp(y) v \right).
\end{equation}
Repeating this and taking the second derivative with respect to $y$ from (\ref{inverse wp expansion}), we find:
\begin{eqnarray}
     \frac{1}{\left( \wp(v)-\wp(y) \right)^3} = \frac{1}{2\wp(y)^3} \left[ \wp'(v-y) + \wp'(v+y) - 2\wp'(y) -\frac{12 \wp'(y)\wp(y)}{\wp(x)-\wp(y)} - \frac{3 \wp'(y)\wp''(y)}{\left( \wp(v)-\wp(y) \right)^2} \right],
\end{eqnarray}
where we have used: $\wp^{(3)}(y) = 12 \wp(y) \wp'(y)$. Remember that the primes denote the derivatives with respect to $y$. Integrating the above expression leads to:
\begin{eqnarray}
     \mathcal{I}_{3} \left( v,y \right) = \int \frac{1}{\left( \wp(v)-\wp(y) \right)^3} = \frac{1}{2\wp(y)^3} \left[ \wp(v+y) - \wp(v-y) - 2\wp'(y) v -12 \wp'(y)\wp(y) \mathcal{I}_1 - 3 \wp'(y)\wp''(y)\mathcal{I}_2 \right].
\end{eqnarray}

\end{document}